\documentstyle[aps,prb,epsf,floats]{revtex}
\begin{document}
\twocolumn[\hsize\textwidth\columnwidth\hsize\csname@twocolumnfalse%
\endcsname
\title{Universal critical temperature for Kosterlitz-Thouless
transitions\\
in bilayer quantum magnets}
\author{Matthias Troyer}
\address{Institute for Solid State Physics, University of Tokyo,\\
Roppongi 7-22-1, Minatoku, Tokyo 106, Japan}
\author{Subir Sachdev}
\address{Department of Physics, Yale University\\
P.O. Box 208120,
New Haven, CT 06520-8120, USA}
\date{\today}
\maketitle
\begin{abstract}
Recent experiments show that double layer quantum Hall systems may have
a
ground state with canted antiferromagnetic order.  In the
experimentally accessible vicinity of a quantum critical point, the
order vanishes at a temperature $T_{KT} = \kappa H$, where $H$ is the
magnetic field and $\kappa$ is a universal number determined by the
interactions and Berry phases of the thermal excitations.  We present
quantum Monte Carlo simulations on a model spin system which support
the universality of $\kappa$ and determine its numerical value. This
allows experimental tests of an intrinsically quantum-mechanical
universal quantity, which is not also a property of a higher
dimensional classical critical point.
\end{abstract}
\pacs{PACS numbers:}
]

Quantum Hall systems offer attractive, tunable laboratories for
investigating zero temperature quantum transitions between states with
different spin magnetizations~\cite{sankar}. Their ground states are
determined almost entirely by the Coulomb interactions between the
electrons, and the typical Coulomb exchange energy is usually much
larger than the Zeeman energy in the external field; as a result, the
spins are often not fully polarized in the direction of the applied
field, and can realize different magnetic configurations which
optimize the Coulomb interactions.

Three separate recent experiments~\cite{pellegrini2,lok,sawada} have
studied magnetic transitions in bilayer quantum Hall systems at total
filling fraction $\nu = 2$. When the layers are well separated, each
layer forms a fully polarized, ferromagnetic state with all states in
the lowest Landau level occupied. The parallel alignment of all spins
is induced mainly by an {\em intralayer ferromagnetic} exchange
interaction~\cite{kallin}, but is also compatible with the Zeeman
coupling which orients the spins in the direction of the magnetic
field. When the layers are closer to each other, there is a
significant {\em interlayer antiferromagnetic} exchange
interaction~\cite{DSZ} which eventually prefers a spin singlet ground
state.  The transition from the fully polarized ferromagnet to the
spin singlet quantum paramagnet could, in principle, be a direct
first-order transition; however, it was
theoretically~\cite{we1} and experimentally~\cite{pinczuk,pellegrini}
found to occur via a softening of the energy of a single spin-flip
excitation in the ferromagnet, suggesting an intermediate phase with
{\em canted} spin ordering~\cite{we1}, bounded by second-order
transitions.  Detailed theoretical predictions of the phase diagram
have been made~\cite{DSZ,ZSD}, and are in good agreement with recent
light scattering experiments~\cite{pellegrini2}.

\begin{figure}
\epsfxsize=3in
\centerline{\epsffile{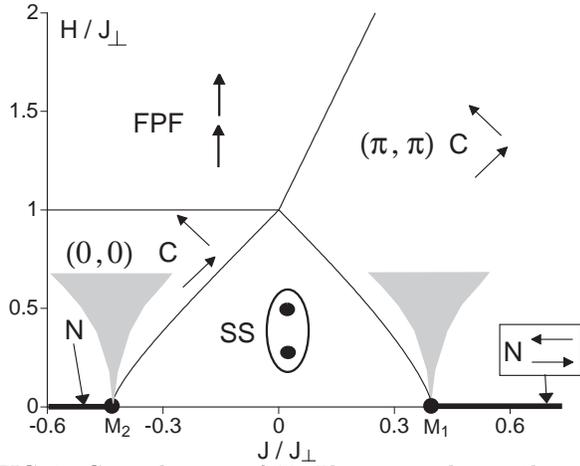}}
\caption{Ground states of ${\cal H}$. The arrows denote the
mean orientations of the spins in the two layers, for the case where
$H$ points vertically upwards; in the SS phase the spins have no
definite orientation.  The in-plane ordering wavevectors of the $x,y$
components of the spins in the C phase are indicated.  The boundaries
of the SS phase approach the points $M_{\mu}$ ($\mu = 1,2$) as
$H/J_{\perp} \sim |w - w_{\mu}|^{\nu}$ where $w \equiv
J/J_{\perp}$, $w_1 = 0.398$ (also determined in
Refs.~\protect\onlinecite{all,gelfand}), $w_2 = -0.435$ (also
determined
in Ref.~\protect\onlinecite{matsu}), and $\nu \approx 0.7$ is the
correlation length exponent of the three dimensional classical
Heisenberg ferromagnet.  There is a Kosterlitz Thouless transition at
non-zero $T$ above the C phases, and the result (\protect\ref{e1})
applies above the shaded regions.}
\label{fig1}
\end{figure}

In this paper, we shall use quantum Monte Carlo simulations to study a
bilayer quantum spin model which has a phase diagram closely related
to that of the $\nu=2$ bilayer quantum Hall system. In particular, all
quantum and nonzero temperature ($T$) phase transitions of the two
models are expected to be in the same universality classes. Our focus
will be on the phase with canted spin ordering. This ordering breaks a
$U(1)$ spin rotation symmetry, and it is expected
that the symmetry will be restored at nonzero $T$ by a
Kosterlitz-Thouless (KT) phase transition at $T=T_{KT}$. In general,
the
value of $T_{KT}$ depends upon microscopic details of the Hamiltonian,
but in the vicinity of a certain quantum critical point (see
discussion below and Fig.~\ref{fig1}) it obeys~\cite{zphys}
\begin{equation}
T_{KT} = \kappa H,
\label{e1}
\end{equation}
where $H$ is the external magnetic field (the electron gyromagnetic
ratio and the Bohr magneton have been absorbed into the definition of
$H$) and $\kappa$ is a non-trivial universal number. It turns out that
our quantum spin model has two separate quantum critical points for
which (\ref{e1}) is expected to be valid.  Our simulations verify that
(\ref{e1}) is indeed obeyed near both critical points; moreover, the
values of $\kappa$ determined at the two points are identical to
within the numerical accuracy, and this supports the claimed
universality of $\kappa$. The same value of $\kappa$ should also apply
to the bilayer quantum Hall systems, and this is a quantitative
theoretical prediction which can be tested in future experiments.

The bilayer quantum spin model has the Hamiltonian
\begin{eqnarray}
{\cal H} = && \sum_{i}\left[ J_{\perp} \hat{\bf S}_{1i} \cdot \hat{\bf
S}_{2i}- {\bf H} \cdot \left(\hat{\bf S}_{1i} + \hat{\bf
S}_{2i}\right) \right] \nonumber \\
&&~~+ \sum_{<ij>} J
\left[ \hat{\bf S}_{1i} \cdot \hat{\bf
S}_{1j} + \hat{\bf S}_{2i} \cdot \hat{\bf
S}_{2j} \right],
\label{e2}
\end{eqnarray}
where $\hat{\bf S}_{ai}$ are quantum spin-1/2 operators in `layers'
$a=1,2$ residing on the sites, $i$, of a two-dimensional square
lattice, ${\bf H} = (0,0,H)$ is the external magnetic field, and
$J_{\perp}$, $J$ are intra- and inter-layer exchange
constants respectively. We will take $J_{\perp} > 0$
antiferromagnetic, but allow $J$ to take either sign.

The model ${\cal H}$ has been studied intensively in recent
years~\cite{all,gelfand,matsu} for the case $H=0$. Using the methods
of Ref.~\onlinecite{annals}, and numerical and exact analytical results
to be discussed below, we obtained the $H \neq 0$ phase diagram shown
in Fig.~\ref{fig1}.
We will now discuss phases in Fig.~\ref{fig1} in turn, and also
indicate the nature of the quantum transitions between them:
\newline
({\em i\/}) Fully Polarized Ferromagnet (FPF): For large enough $H$,
the exact ground state is simply the state with all spins up {\em
i.e.} in the eigenstate of $\hat{S}_{aiz}$ with eigenvalue $1/2$. The
first excited state is a single spin-flip, and its excitation energy
can be determined exactly: at momentum $\vec{k} = (k_x, k_y)$ it is
$\epsilon_{\vec{k}} = H - J_{\perp} - J ( 2 -
\cos k_x - \cos k_y )$. For $J > 0$ ($J <
0$) this has a minimum at $\vec{k} = (\pi, \pi)$ ($\vec{k} =
(0,0)$). Stability of the FPF state requires that $\epsilon_{\vec{k}} =
\geq
0$, and the point where the minimum energy first vanishes
exactly determines the FPF phase boundary shown in Fig.~\ref{fig1}.
The single spin-flips Bose condense at this boundary, leading to a
phase with canted spin order to be described below. This
transition is in the universality class of the dilute Bose gas
quantum critical point~\cite{annals} with dynamic exponent $z=2$.
\newline
({\em ii\/}) Spin Singlet (SS): This is the spin singlet quantum
paramagnet with no broken symmetries and a gap to all excitations.
The exact ground state is known only for $J = 0$, when the
system decouples into pairs of spins antiferromagnetically coupled by
$J_{\perp}$, which therefore form a spin singlet valence bond (see
Fig.~\ref{fig1}). The ground state for $J \neq 0$ is
adiabatically connected to this decoupled state, and its wavefunction
can be determined~\cite{gelfand} in an expansion in powers of
$J/J_{\perp}$.  For $H=0$, the lowest excited state is
triplet particle, again adiabatically connected to the $J
= 0$ limit, whose dispersion has been computed~\cite{gelfand} in a
series in $J/J_{\perp}$. Turning on a nonzero $H$ leads to
no change in the wavefunctions, but the energy of the particle changes
by $\epsilon_{\vec{k}} \rightarrow \epsilon_{\vec{k}} - m H$, with
$m=1,0,-1$ the $S_z$ quantum number. As for the FPF phase, the
boundary of the SS phase is the line where the minimum excitation
energy vanishes. For $J > 0$, the boundary is at
$H/J_{\perp} =1 - 2 w - (3/2) w^3 - (3/2) w^4 + {\cal O} (w^5)$, with
$w \equiv J/J_{\perp}$, while for $J < 0$ it
is at $H/J_{\perp}= 1 + 2 w - (7/4) w^4 + {\cal O}(w^5)$. For $H > 0$,
there is a Bose condensation of the particle at this line, which is in
the same universality class as the boundary of the FPF phase, and also
leads to the same canted phase. Precisely at $H=0$ (the points $M_1$,
$M_2$ in Fig.~\ref{fig1}) the nature of the transition is different,
and will be discussed below.
\newline
({\em iii\/}) Canted (C): ${\cal H}$ has a symmetry of rotations about
the axis ($z$) of the applied field, and this is broken in this
phase. The spin operators have the expectation values $\langle
\hat{S}_{1z} \rangle = \langle \hat{S}_{2z} \rangle \neq 0$ and
$\langle \hat{S}_{1x,y} \rangle = -\langle \hat{S}_{2x,y} \rangle \neq
0$. The $z$ expectation values are independent of $i$, while the $x,y$
expectation values have staggered (uniform) arrangement on the two
sublattices with each layer for $J > 0$ ($J <
0$). Associated with the broken symmetry, there is linearly dispersing
Goldstone mode of spin-wave excitations corresponding to slow
rotations of the order parameter in the $x,y$ plane.
\newline
({\em iv\/}) N\'{e}el (N): This is reached in the $H=0$ limit of the
$C$ phase, when $\langle \hat{S}_{az} \rangle = 0$. Now ${\cal H}$ has
the full $SU(2)$ spin rotation symmetry, and so the spin expectation
values in the $x,y$ plane can actually point along any direction in
spin space.

The vicinities of the critical points $M_{1,2}$ are of particular
interest in this paper. Here the system is expected~\cite{DSZ,ZSD} to
be described by a continuum quantum field theory which can be
expressed in terms of a unit length field ${\bf n} (x, \tau)$ ($\tau$
is imaginary time). For $M_1$, ${\bf n} \propto \sum_{i \in {\cal
N}_x} (-1)^{i_x + i_y} (\hat{\bf S}_{1i} - \hat{\bf S}_{2i})$ and for
$M_2$, ${\bf n} \propto \sum_{i \in {\cal N}_x} (\hat{\bf S}_{1i} -
\hat{\bf S}_{2i})$, where ${\cal N}_x$ is an averaging neighborhood of
$x$. The field theory has the action
\begin{equation}
{\cal S} = \frac{c}{2g} \int d^2 x  d \tau
\left[ (\nabla_x {\bf n} )^2 + \frac{1}{c^2} \left( \frac{\partial
{\bf n}}{\partial \tau}
+ i {\bf H} \times {\bf n} \right)^2 \right],
\label{e3}
\end{equation}
with $c$ a velocity and $g$ a coupling constant which tunes the value
of $J/J_{\perp}$. At $H=0$ this theory can be
reinterpreted as the real partition function of a three-dimensional
classical Heisenberg ferromagnet at non-zero temperature. However no
such classical interpretation is possible for $H \neq 0$: notice then
that the action is {\em complex} (even in imaginary time), as it
includes the Berry phase of the precession of the ${\bf n}$ quanta
about the applied field. Applying a field at the scale-invariant
critical points $M_{1,2}$, we can conclude that all characteristic
temperatures will be determined by $H$. Scaling
arguments~\cite{zphys}, which rely on the fact that ${\bf H}$ appears
in ${\cal S}$ as the time component of a $O(3)$ non-Abelian gauge
field (which is in turn related to the fact that ${\bf H}$ couples to
the conserved total spin), show that $H$ scales as an inverse time; as
temperatures also scales as inverse time, we are then lead to
(\ref{e1}) for the critical temperature at which the field-induced
canted order will disappear.

We turn now to our quantum Monte Carlo results, obtained using the
powerful loop algorithm\cite{loops} successfully used in recent
studies of quantum Heisenberg\cite{QHB} and XY models\cite{QXY}. All
weights are positive in this basis if we apply the field along the
axis of quantization, and the Berry phases have been transformed into
the quantization of the spins on the world lines\cite{foot}. The loop
algorithm slows down severely for $H \neq 0$ \cite{wormchains}, as a
loop that changes the magnetization by $\Delta M$ picks up a flipping
probability $\exp( H\Delta M/T)$; this leads to an exponential
increase of the autocorrelation times from $\tau_{\rm int}<1$ to
$\tau_{\rm int}\propto\exp( H/T)$. Fortunately it is sufficient to
simulate at not too low $T$, where $ H/T <4$ and $\tau_{\rm int}<100$,
but our system sizes, $L$, are not as large as in earlier $H=0$
studies\cite{QHB,QXY}.

\begin{figure}
\epsfysize=2.9in
\centerline{\epsffile{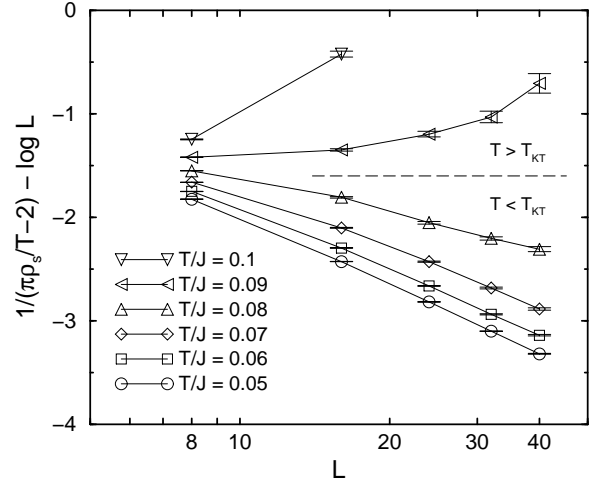}}
\caption{Scaling plot of $1/(\pi\rho_s/T-2) -\log L$ versus $L$. This
quantity, which diverges to $+\infty$ for $T>T_{KT}$ and to $-\infty$
for $T<T_{KT}$ allows a reliable estimate of $T_{KT}$.}
\label{fig:scaling}
\end{figure}

We obtained $T_{KT}$ following the method of
Harada and Kawashima\cite{QXY} for the quantum XY model.
The spin stiffness, $\rho_s$, of the $U(1)$ ordering in the $x-y$
plane was obtained from $\rho_s = T \left\langle W_x^2 + W_y^2
\right\rangle/2$, where $W_{x,y}$ are the total winding
numbers in the two directions.
The improved estimators in
Ref. \onlinecite{QXY} have to be modified, since the flipping
probabilities of loops are no longer all equal, and a multi-loop
algorithm is necessary. For $L=\infty$, $\rho_s$ is finite for
$T< T_{KT}$, continuously decreases to the universal value
$\rho_s(T=T_{KT}) = (2/\pi) T_{KT}$ at $T_{KT}$ and is zero for
all $T>T_{KT}$.
However, for $L$ finite, $\rho_s$ is nonzero at all $T$, and is
expected to converge to the $L=\infty$ limit with
 the finite size scaling
form at $T=T_{KT}$ \cite{QXY,Weber}:
\begin{equation}
\pi \rho_s / 2 T = 1 + [2 \log (L/L_0 (T))]^{-1}.
\label{eq:fss}
\end{equation}
Hence, good estimates for $T_{KT}$ can be obtained by plotting
$1/(\pi\rho_s/T-2) -\log L$ as a function of $L$.  As $L$ is increased
this quantity converges to the constant $(-\log L_0)$ at $T_{KT}$, and
diverges to $\pm\infty$ for $T>T_{KT}$ and $T<T_{KT}$ respectively.
Our data (a representative example is shown in
Fig.~\ref{fig:scaling}) are clearly consistent with these expectations.
Note that we obtained $L_0\approx 5$ to $10$, compared to $L_0=0.23$
in the $XY$ model\cite{QXY}; so we cannot use the more elaborate
fitting techniques used in Ref. \onlinecite{QXY}, as we are not
deep enough in the asymptotic scaling regime.
\begin{figure}
\epsfxsize=3in
\centerline{\epsffile{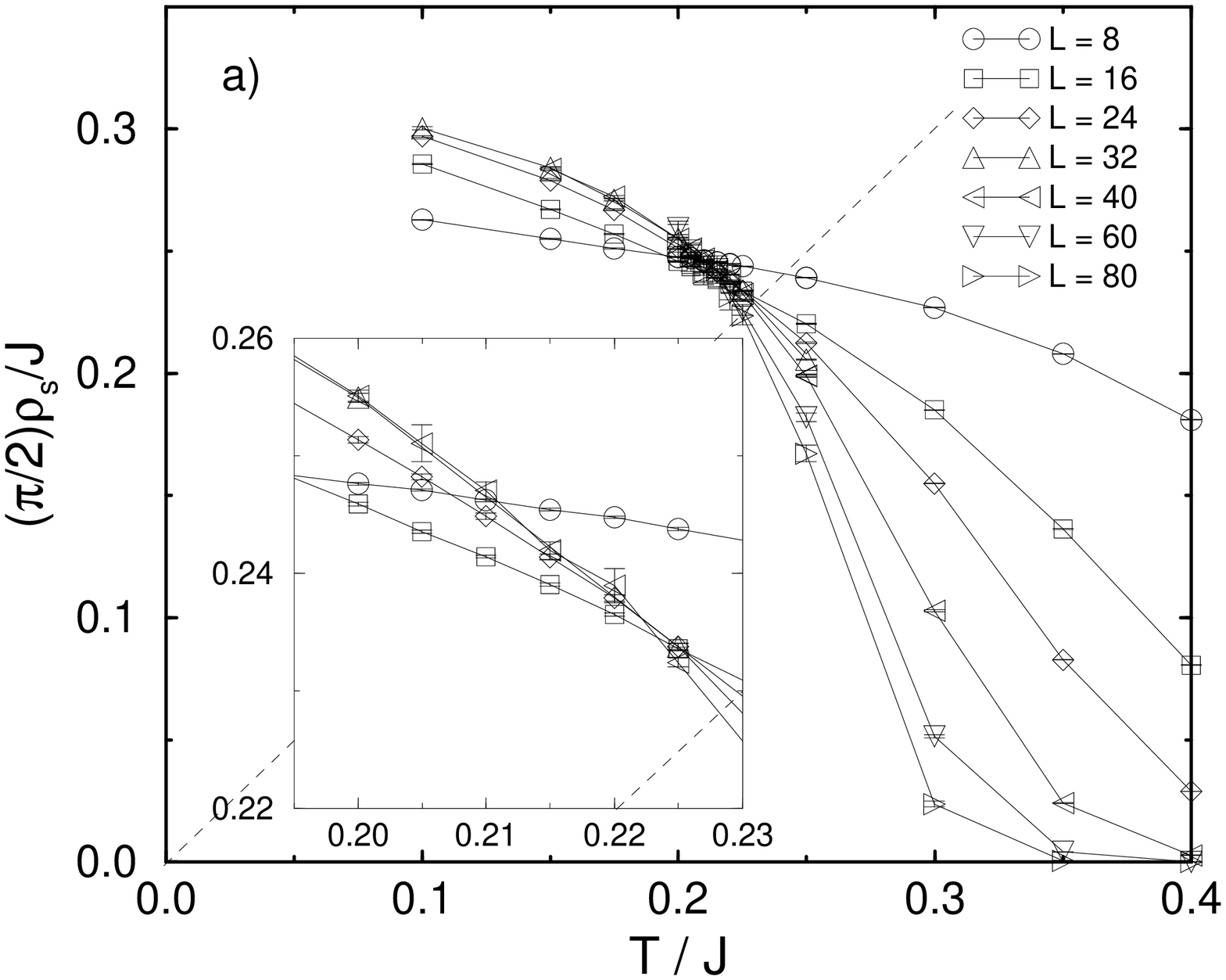}}
\epsfxsize=3in
\centerline{\epsffile{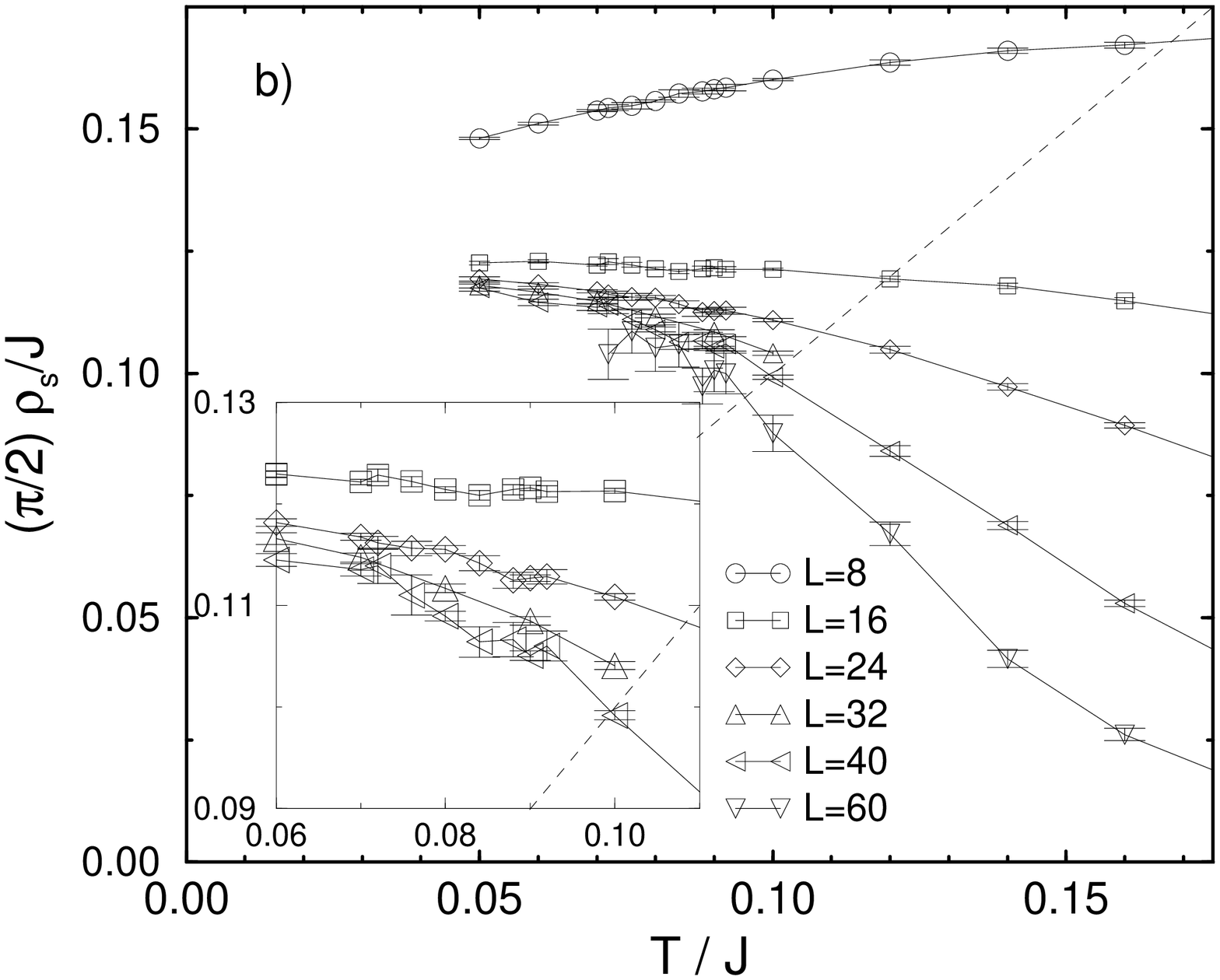}}
\caption{Spin stiffness $\rho_s$ in a magnetic field $H=0.2J$ of
${\cal H}$
for (a) the single layer model ($J_{\perp} = 0$) and (b) vertically
above the critical point $M_1$ of Fig.~\protect\ref{fig1}, as a
function of $T$ for various $L$.
The dashed lines are the lines of
universal values at $T_{KT}$: $(\pi/2) \rho_s(T_{KT}) = T_{KT}$.
}
\label{fig2}
\end{figure}

First, we checked for a KT transition for the decoupled single-layer
case ($J_{\perp} = 0$).  Our results for $\rho_s$ ({\em per\/} layer)
are shown in Fig. \ref{fig2}a. We find KT transitions at $T_{KT}/J =
0.190(5)$ for $H/J=0.1$ and at $T_{KT}/J = 0.215(5)$ for
$H/J=0.2$.
Surprisingly however, anomalous non-monotonic finite size scaling
behavior was found below $T_{KT}$. As can be seen in Fig.~\ref{fig2}a,
when increasing $L$, first $\rho_s$ decreases, then increases again,
and is asymptotically expected to decrease again, like in the XY
model. A similar anomalous scaling was found in this model by Lavalle
{\it et al.} \cite{lavalle} for the ground state energy in subspaces of
nonzero spin $S\propto L^2$.

\begin{figure}
\epsfysize=3in
\centerline{\epsffile{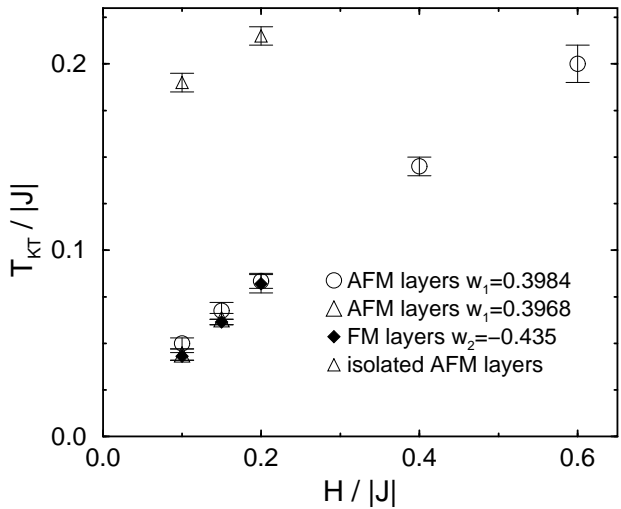}}
\caption{The critical temperature $T_{KT}$ as a function of field $H$
for
the square lattice (isolated layers) and for the antiferromagnetic
(AFM) and ferromagnetic (FM) bilayer models close to both critical
points.}
\label{fig3}
\end{figure}

We now turn to fields above the critical points
$M_1$ and $M_2$. We performed simulations in fields
$H/J=0.1$, $0.15$, $0.2$, $0.4$, $0.6$ and $1$. Here no anomalous
finite size scaling is observed, as can be seen in Fig.~\ref{fig2}b
(the $\rho_s$ now is the total value, not per layer).
We can thus confidently use the finite size scaling analysis
to determine $T_{KT}$ as a function of $H$, and show the
results in Fig.~\ref{fig3}.

We fit the results of Fig.~\ref{fig3} $H/|J|\le0.2$ to $T_{KT} = a (
\delta ) +
\kappa H$, where the offset $a(\delta)$ accounts for the error in our
determination of the positions of $M_{1,2}$.
We obtained for the slope $\kappa=0.35(7)$ at $M_1$ and
$\kappa=0.37(5)$ at $M_2$. These numbers agree very well. The error
bars are our estimate of strict upper and lower bounds, and not
one-sigma confidence intervals. We obtained $a ( \delta ) \approx
0.01J$, comparable with what the $H=0$, $T=0$ $\rho_s$ could be,
given
the uncertainty of about 0.4\% in the determination of the critical
points (compare the results for slightly different coupling ratios
in Fig.~\ref{fig3}).

Corrections to scaling from not being in the continuum limit appear at
$H \sim J$ and lead to $T_{KT} = \kappa H ( 1 - bH/|J|)$: these are
smaller and were ignored in above fits for $H/|J|\le0.2$. Using above
estimates for $\kappa$ and $a(\delta)$ and a small correction
$b\approx0.3$ we can however fit all our results up to $H=0.6|J|$.
This correction slightly increases our final combined estimate:
\begin{equation}
\kappa\approx0.38\pm0.06;
\end{equation}
we also recall the leading result in the $\epsilon=3-d$
expansion~\cite{DSZ,ZSD}: $\kappa = \sqrt{33/10\pi^2 \epsilon}
\approx 0.58$.

To conclude, we have obtained a numerical estimate for the
universal temperature of a KT transition in the vicinity of a
quantum critical point.
This universal quantity relies on an
underlying interacting quantum field theory with complex Berry
phases in $d=2$, and experiments to measure it in bilayer
quantum Hall systems will be of considerable interest.
Current measurements~\cite{pellegrini2,pellegrini}
of $T_{KT}$ are in the vicinity of (\ref{e1}), but more detailed
measurements in the shaded region of
Fig~\ref{fig1} are necessary, possibly by pressure tuning of the
$g$-factor~\cite{nicholas}.

We thank S. Das Sarma and M. Gelfand for discussions.
This research was supported by NSF Grant No DMR 96--23181. The QMC
calculations were performed on the Hitachi SR2201 massively parallel
computer of the University of Tokyo, using a parallelizing Monte Carlo
library in C++ developed by one us\cite{ALEA}.


\begin{references}
\bibitem{foot} For $H\neq 0$, any basis will have either complex Berry phases or
quantized spins. In contrast, for $H=0$, ${\cal S}$ can be
represented by a continuous spin classical ferromagnet in $d=3$.
\end{references}
\end{document}